**Great Short History of Microbiology Development as a Science**

Daniil S. Gerassimov

Department of Natural Mathematical Sciences, Kozybayev University

Microbiology

Prof. Zhadan

March 25, 2025

**Table of Contents**

## History of Microbiology Development as a Science

**Introduction**

The study of microorganisms, or microbiology, has demonstrated significant development since its inception and is currently a key field of biological sciences that has a huge impact on modern society and scientific research. Over the centuries, this discipline has undergone significant changes, shaping our understanding of infectious diseases and food safety. Starting from the simplest observations of microscopic organisms such as bacteria, viruses, fungi and protozoa, and ending with modern molecular and genomic research methods. This article describes a brief historical path of microbiology development. The heuristic, morphological, physiological, immunological, and molecular genetic stages are the main periods into which the development of this science is traditionally divided, despite the lack of full-fledged and precise boundaries between them.

**Heuristic period**

The heuristic period in the development of microbiology covers a significant time period, from ancient times to the middle of the 19th century. The exact time frame may vary depending on how researchers define the beginning and end of this period, but it usually includes ancient times. From the first assumptions about the nature of diseases and fermentation made by ancient Greek and Roman philosophers to the Middle Ages, and more precisely to the middle of the 19th century. The period ends with the beginning of active experimental research and the creation of a microscope, when microbiology passes into the physiological period (Gest, 2004).

The heuristic period in the development of microbiology is characterized by assumptions about the causes of infectious diseases when microorganisms were not yet known in the period of antiquity. Despite this, people have already used bacterial waste products, for example, in alcoholic, lactic acid, and acetic acid fermentation in the production of wine, bread, beer, and cheese (Santer, 2010).

Many scientists have suggested that fermentation, putrefaction, and diseases are caused by invisible creatures. The ancient Greek physician Hippocrates believed that contagious diseases arise from invisible substances, "miasma," formed in swampy places. For his contribution to medicine, Hippocrates is called the "father of medicine " (Gest, 2004).

The ancient Roman poet and philosopher Titus Lucretius explained contagious diseases by the presence of "invisible seeds" specific to each infection. The Spartan military commander and philosopher Thucydides assumed the existence of a "living contagium" (contagium animatum) that causes infectious diseases, and even formulated the idea of immunity to repeated disease. The ancient Roman scientist Marcus Terentius Varro also believed that epidemics were caused by "tiny animals" (animalcula quaedam minuta) (Littman, 2009).

| Scientist | Lifespan | Contribution to the Study of Infectious Diseases |
|---|---|---|
| Marcus Terentius Varro | 116-27 BC | Suggested the existence of "minute animals" causing epidemics. |
| Rhazes (Abu Bakr Muhammad ibn Zakariya) | 865-925 AD | Hypothesized the infectious nature of some diseases, provided classical descriptions of smallpox and measles, recommended smallpox inoculation. |
| Avicenna (Abu Ali al-Husayn ibn Sina) | 980-1037 AD | Proposed that diseases are caused by minute beings, noted the contagiousness of smallpox, distinguished between cholera and plague, described leprosy. |
| Maimonides (Abu Amram Musa ibn Maymun) | 1135-1204 AD | Established preventive medicine as a separate field, recommended special attention to the care of those recovering from infectious diseases. |
| Theophrastus Paracelsus | 1493-1541 AD | Hypothesized that contagious diseases are caused by living beings, developed methods and remedies for treating contagious diseases, introduced mercury, sulfur, and iron preparations into medical practice. |
| Girolamo Fracastoro | 1476-1553 AD | Substantiated the theory that contagious diseases are caused by "living contagions," suggested isolating the sick and fumigating premises with juniper, introduced the term "infection" into medicine. |

Table 1. Theories and scientists The Heuristic Period

Thus, the heuristic period covers millennia, during which knowledge about microorganisms was accumulated on the basis of observations and assumptions, without using accurate research methods. Despite the lack of evidence, he laid the foundation for further research. The proposed ideas stimulated the development of science and the formation of ideas about the microcosm (Gest, 2004).

**Morphological period**

The morphological period in the development of microbiology covers the time from the invention of the microscope to the beginning of active physiological studies of microorganisms. Usually, the beginning of this period is called the middle of the 17th century to 18-19 (Wainwright, 2003). Thus, the morphological period spans approximately two centuries, during which the main focus of research was the study of the shape and structure of microorganisms using a microscope.

The foundations of microbiology as a science were laid in the 17th century thanks to the pioneering work on the invention of the microscope. With the use of the microscope, a new stage in the development of microbiology began—the morphological period—the period of discovery of the world of microbes. For the first time, a magnifying optical device of this kind was invented by Galileo in 1612, which was a telescope with a small magnification, which is why the device was only suitable for studying insects (Collier, 2008). The Jansens in the Netherlands have designed the first microscope suitable for detecting large microbes, magnifying it by 32 times. Now the device consisted of two convex lenses inside one tube; that is, it was the prototype of a modern telescope rather than a microscope. Nevertheless, Athanasius Kircher, an alchemist and a Roman teacher, discovered "worms" in rotting foods and the blood of plague patients using the Jansen microscope. He believed that the living beings he observed originated from inanimate organic compounds.

Anthony van Leeuwenhoek became most famous during this period of development of microbiology. In 1673, he created a microscope that magnifies 150-300 times. The Leeuwenhoek microscope was a biconvex lens with a very short focal length, so when working, the microscope had to be brought very close to the eyes. Due to this, he described ciliates, giardia, erythrocytes, and bacteria, calling them "animalcules" (Ford, 1991).

The greatest contribution to the development of the microscope was made by Robert Hooke, who improved the microscope by creating a complex lens system. He created a prototype based on the principle of which seizes is also used in light microscopes, namely, a lens and an eyepiece were created, and magnification and resolution were recognized as the most important characteristics of the microscope. At the same time, Robert Hooke's work, in particular his publication Micrography, published in 1665, provided additional evidence for the existence of microscopic life and helped establish microscopy as a scientific tool (Gest, 2004).

To prove the role of microbes in pathology, animal experiments and self-disinfection were conducted. So, Danilo Samoilovich Sushchinsky injected himself with the contents of a bubonic plague patient, after which he contracted the plague as a result of self-infection in a mild form. Similar methods were used in the future. In the 18th and 20th centuries, many infectious agents were discovered, but the causative agents of measles, polio and influenza remained unknown (Blevins and Bronze, 2010).

In 1898, Martinus Beyerink was able to isolate the causative agent of tobacco mosaic and named it a virus using marble filters. In the following years, viruses infecting humans, animals, and bacteria (bacteriophages) were discovered. Also, as early as the 20th century, descriptions of microorganisms continued, for example, bacteriophages were described by Frederick Tuort in 1915 and Felix Darel in 1917 (Carter, 1988). Darel proposed using bacteriophages to treat bacterial diseases and successfully cured the first patient.

At this stage in the development of the study of microorganisms, the study and discovery of new types and causes of diseases continued. Virology was formed in the first half of the 20th century. The discovery of new types of microorganisms continues. In recent decades, HIV, pathogens of hemorrhagic fevers, legionnaires' disease, and SARS have been discovered. Many microorganisms have acquired new properties and have become pathogenic to humans. Thanks to vaccination, smallpox was eliminated.

The morphological period spanning the 17th and 19th centuries became the foundation for modern microbiology, marking the transition from guesswork to direct observation of the microcosm. The invention of the microscope made it possible for the first time to see and describe a variety of microorganisms. This period was characterized by the active accumulation of knowledge about the shapes, sizes and structure of microbes, which played a crucial role in understanding their role in nature and pathology. Although the physiological and biochemical aspects of microbial life remained poorly understood, morphological studies laid the foundation for subsequent discoveries, including establishing a link between microbes and infectious diseases (Worboys, 2007).

However, despite these early observations, the systematic study of microorganisms remained limited due to both technological limitations and the prevailing theoretical foundations of the time. The debate about the spontaneous generation of microorganisms, which continued until the 19th century, had a significant impact on early microbiology. At the center of these debates was the question of whether microorganisms could arise spontaneously from inanimate matter or whether their formation required pre-existing life (Farley, 1977).

### Physiological Period

The physiological period of microbiology development covers the time period from the middle of the 19th century to the beginning of the 20th century. This period is characterized by an intensive study of the physiological and biochemical properties of microorganisms, as well as their role in various processes, including diseases.

After the discovery of microbes, scientists' close attention was drawn to the issues of structure, biological properties, and vital activity of pathogens of infectious diseases. Thus, the physiological period of microbiology (the middle of the 19th century) began with the study of the structure and properties of pathogens. Louis Pasteur and Robert Koch are considered the founders of the physiological period of microbiology (Blevins & Bronze, 2010).

Pasteur discovered the nature of fermentation and anaerobiosis, refuted the theory of spontaneous generation, and justified sterilization and vaccination, which allowed him to become the founder of stereochemistry, microbiology, immunology, biotechnology, and disinfection (Porter, 1976).

In 1857, Pasteur established that fermentation is caused by bacteria, and in 1860 he proved the impossibility of self-generation of microbes (Dubos, 1998). He experimentally proved that germs from the air multiply in a sterile boiled broth in an open flask. But if the sterile broth is in a flask that communicates with the air through a spiral curved glass tube, the broth will remain sterile for a long time, since bacteria from the air will be deposited in water droplets in the curved part of the tube and will not enter the broth. By these experiments, L. Pasteur refuted the theory of spontaneous generation of living organisms that existed at that time. Pasteur established that certain microorganisms are responsible for certain chemical transformations, and this concept later became central to industrial microbiology.

Thus, in 1865, L. Pasteur established that microorganisms entering wine or beer from the external environment cause spoilage of these products, while others are responsible for fermentation and the appearance of alcohol in foreign juice. He suggested heating such products at a temperature of 60-80°C, which was sufficient to destroy harmful bacteria but did not affect the quality of the products. Developed a method in honor of L. Pasteur, which became known as "pasteurization." To sterilize some products, L. Pasteur proposed heating them to 120°C in a steam boiler, which was called an "autoclave.". Microorganisms have also become widely used in fermentation, enzyme production, and biotechnology. An example is yeast (Saccharomyces cerevisiae), which plays an important role in brewing and baking Porter, 1976).

On April 30, 1878, in his report to the French Academy of Sciences, L. Pasteur pointed out that 26 microorganisms are the cause of infectious diseases of humans and animals. This day is considered the birthday of medical microbiology as a science (Vallery-Radot, (1985). It is believed that it was Pasteur who developed the principles of vaccination, although scarification and variolation were used before him. In China, India, and Iraq, to prevent smallpox, children had incisions (scratches) applied to their skin and pus from the pustules of sick people rubbed into them (Dubos, 1998). This method was called scarification. After such scarification, the disease developed, which proceeded relatively easily and left behind immunity to subsequent infection with the same pathogen. Another method of protection was also used for smallpox, the so-called variolation—the introduction of dried smallpox crusts into the nostrils, applying them to incisions made on the skin.

Already in 1796, Edward Jenner created a vaccine against smallpox. He used a liquid taken from a vial from the hand of a woman who had contracted cowpox while milking a cow. Two months after that, E. Jenner injected it into the vaccinated boy. The boy did not get smallpox. After 5 months, the boy was re-injected with material taken from a smallpox patient. The disease did not occur. In this way, the possibility of artificially creating immunity to smallpox was established (Blevins & Bronze, 2010).

However, it was Pasteur who first proposed the method of attenuation (attenuation) of pathogenic strains of microorganisms by long-term cultivation on nutrient media or by long-term passaging through the body of laboratory animals. He conducted an experiment with a live culture of the causative agent of chicken cholera for a long time in a test tube without reseeding. After the introduction of such an "old" culture, chickens did not get sick with cholera. But the surprising thing was that the introduction of fresh culture to these chickens also did not lead to the development of the disease. Using this method of attenuation, L. Pasteur in 1881 created vaccines against anthrax (Schwartz, 2001).

Robert Koch (1843-1910) developed methodological approaches that transformed microbiology into a rigorous scientific discipline. His work on anthrax and tuberculosis led to the formulation of the "Koch postulates," which established criteria for determining whether a particular microorganism causes a particular disease (Blevins & Bronze, 2010). Koch also pioneered methods for cultivating bacteria on solid nutrient media, allowing for the isolation of pure cultures and facilitating the identification of pathogenic organisms. The work of Martinus Beyerink (1851-1931) and Sergei Vinogradsky (1856-1953) took microbiology beyond pathogenic organisms to include environmental microbes. Their research laid the foundations for the concept of microbial ecology and the role of microorganisms in the biogeochemical cycle. During this period, the identification of viruses as separate infectious agents also began, starting with the studies of tobacco mosaic disease conducted by Dmitry Ivanovsky in 1892 and the subsequent characterization of tobacco mosaic virus by Martinus Beyerink as a "contagium of living fluid."

The physiological period was important for the development of modern microbiology, including the study of the processes of nutrition, respiration, growth, and reproduction of bacteria. It has been revealed and proven that the vital activity of microorganisms affects various processes, including fermentation, putrefaction, and diseases (Santer, 2010).

The theory of the spontaneous generation of life has been completely refuted, proving that microorganisms arise only from existing microorganisms. Also during this period, nutrient media for growing microorganisms in the laboratory were developed, which made it possible to study their properties and behavior (Blevins & Bronze, 2010). All this laid the foundation for the development of modern microbiology, immunology, and medicine.

**Immunological period**

The immunological period of microbiology (the end of the 19th and the first half of the 20th centuries) began with the development of methods of protection against pathogenic microbes. This period is closely related to Mechnikov and Ehrlich. These scientists can rightfully be called the founders of immunology, since Mechnikov proposed a cellular (phagocytic) theory of immunity, and P. Ehrlich developed a humoral theory of immunity (Gest, 2004).

During this period, scientists began to actively develop ways to protect themselves from pathogenic microbes that cause infectious diseases in humans and animals. In particular, Mechnikov was the first to develop the doctrine of phagocytes and phagocytosis. In 1883, he established that phagocytes, amoeboid motile cells that "devour" pathogens, perform the main function of protecting the body from infections. I.I. Mechnikov noted that phagocytosis is observed in all animals. Mechnikov's opponent was Paul Ehrlich, who proposed a humoral theory of immunity. P. Ehrlich believed that in response to the introduction of microbes or their toxins into the body, specific protective substances are produced – antibodies. In this regard, he attached primary importance to antibodies (humoral factors) in the development of immunity. However, further development of immunology has shown the unity of cellular and humoral immunity factors. P. Ehrlich and I.I. Mechnikov were awarded the Nobel Prize in 1908 for developing the theory of immunity (Littman, 2009).

The first half of the 20th century was marked by the rapid development of immunology. In 1898, Borde showed that antibodies are formed against any foreign protein (antigen) and not just to bacteria or their toxins (Blevins & Bronze, 2010). In 1904, Pirke and Richet discovered anaphylaxis (hypersensitivity in response to the introduction of a foreign protein into the body).

Thus, at the beginning of the 20th century, a new science arose - immunology, which was divided into two areas: - infectious immunology; - non-infectious immunology. Until the middle of the twentieth century . Infectious immunology was mainly developing. Vaccines and therapeutic serums were created against the most common infections (plague, tuberculosis, typhoid fever, diphtheria, etc.). However, theoretical immunology remained in its infancy (Gest, 2004) .

A new stage in the development of immunology began in the late 1940s, when Frank MacFarlane Burnet combined these areas into the so-called "new immunology." In 1949, F.M. Burnet proposed a clonal selection theory of immunity, according to which lymphocytes undergo selection in the embryonic period. Those of them that show aggression towards "their" antigens are destroyed. The remaining lymphocytes react only with foreign antigens, thus formulating and opening the immunological memory (Littman, 2009).

In 1953, Peter Medawar and Milan Hasek experimentally confirmed the theory and found that animals that were injected with foreign antigens in the embryonic period perceived them as "their own" after birth.). In the second half of the twentieth century, it was found that cells of the immune system are able to "communicate" with each other using specialized mediators - cytokines. Cytokine signals are perceived only by those cells on the 38th surface of which there are specific receptors. Currently, the development of immunology continues (Gest, 2004). The work of Alexander Fleming, Howard Florey, and Ernst Chain on the discovery and development of penicillin in the 1920s and 1940s ushered in the era of antibiotics, revolutionizing medicine and stimulating research into secondary metabolites of microorganisms that significantly reduced mortality from bacterial infections.

**Molecular Period – Present**

The molecular genetic era of microbiology development began in the second half of the 20th century and continues to this day. It was marked by James Watson and Francis Crick, who created the structure and three-dimensional model of the DNA molecule, and Avery, McLeod, and McCarthy, who discovered the role of DNA in heredity (1944). Since then, biotechnology, gene and protein engineering, and molecular biology have become widespread (Handelsman, 2004).

New methods have made it possible to study the pathogenicity factors of microorganisms, the structure of antigens and antibodies, as well as the genome structure of numerous bacteria and viruses. It is now possible to create synthetic recombinant DNA molecules and breed bacterial strains with new characteristics by decoding the genes of viruses and bacteria (Marx, 2013).

While sophisticated light microscopy techniques such as confocal and fluorescence microscopy have made it easier to study microorganisms in vivo and in real time, the development of electron microscopy in the 1930s made it possible to visualize the ultrastructure of bacteria and viruses (Gest, 2004).

The identification and characterization of microorganisms has been improved using molecular techniques such as DNA sequencing, gene expression analysis, and polymerase chain reaction (PCR) (Handelsman, 2004).

Since new methods make it possible to study huge amounts of genetic data and model metabolic networks of interaction with the environment, bioinformatics and computational biology have become the most important elements of modern microbiology (Marx, 2013).

One example of such work using genetics is the work of Paul Berg, who created recombinant DNA in vitro in 1972. This DNA was made up of fragments of several bacterial and viral nucleic acids. After decoding the E. coli genome, he was able to transfer individual genes from one cell to another and create artificially engineered genes. Thus, microorganisms became tools of genetic engineering and biotechnology in the 1970s, when Paul Berg, Herbert Boyer, and Stanley Cohen developed recombinant DNA technology (Hughes, 2001). It is impossible not to mention the CRISPR-Cas9 system, found in the protective mechanisms of bacteria, has become a revolutionary tool for gene editing, which finds application in medicine, agriculture and synthetic biology.

As science has developed, microbiology has been divided into sections that focus on various aspects of microbial life and their use. The study of harmful microorganisms and how they interact with the human body has led to the emergence of the independent science of medical microbiology. One of the notable achievements in this field is the development of vaccines and methods for the detection and treatment of infectious diseases (Poupard et al., 2010).

Microorganisms are used in industrial microbiology to create useful substances such as food additives, enzymes, antibiotics, and biofuels. Using genetically modified microbes, the fermentation industry based on outdated technologies has turned into a complex biotechnology sector (Handelsman, 2004).

Environmental microbiology studies how microorganisms function in the natural environment and how they can be used for waste disposal and bioremediation. Research in this area has revealed the most important roles that bacteria play in soil fertility, nutrient cycling, and ecosystem health (Gest, 2004).

Understanding the genetic basis of microbial behavior characteristics is the main goal of molecular microbiology and microbiological genetics. These disciplines have made fundamental contributions to biology through the adaptive responses of microorganisms, communication related to the sense of bacterial quorum, and horizontal gene transfer (Handelsman, 2004). Currently, a number of biologically active substances (BAS) are being produced using genetic engineering methods, which are used as preventive, therapeutic and diagnostic agents.

### Modern Challenges in Microbiology

The main ones are antimicrobial resistance, new infectious diseases, and environmental changes - these are just some of the major challenges facing modern microbiology. The COVID-19 pandemic caused by the SARS-CoV-2 virus has highlighted the continuing danger of infections and the importance of timely diagnosis and vaccine development. Antimicrobial resistance is one of the most pressing public health problems requiring innovative methods of finding antibiotics and treatment. Innovative screening methods and previously unknown microbiological sources are increasingly being used in the search for new antibacterial chemicals (Gest, 2004).

### Conclusion

Microbiology has undergone significant changes over time, from basic observations to complex molecular knowledge. The understanding of the world, industry, and human healthcare have all changed as a result of technological advances, theoretical discoveries, and practical applications. Starting from Van Leeuwenhoek's early microscopes and ending with modern genomics. It faces both opportunities and obstacles when using new technologies and integrating with other disciplines. Our knowledge of the diversity and capabilities of microorganisms is still insufficient, as evidenced by the constant discovery of new species and their role. Future developments in the field of microbiology can help solve global problems such as the emergence of new diseases. Modern microbiologists can evaluate the origins of this discipline in order to understand possible ways of its development in the future, about its history.